\documentclass[apj]{emulateapj}\usepackage{apjfonts}\epsscale{1.05}\newcommand{\resizedraft}{}
\usepackage{natbib}
\usepackage{textcomp}
\usepackage{aas-macros}
\usepackage{amsmath}
\usepackage{amssymb}
\bibpunct[, ]{(}{)}{;}{a}{}{,}
\newcommand{\D}{\displaystyle}

\newcommand{\tabover}[2]{$\D\begin{array}{c}%
                             {#1}\\{#2}
                         \D\end{array}%
$}
\newcommand{\tabunit}[2]{\begin{array}{c}%
                             {#1}\\{(#2)}
                         \end{array}%
}
\newcommand{\tabhead}[1]{\caption{#1}}
\newcommand{\pct}{\,\%}
\newcommand{\grad}{\mbox{\textdegree}}
\newcommand{\dez}[1]{{}\times10^{#1}\,}
\newcommand{\fracdisp}[2]{\frac{\displaystyle #1}{\displaystyle #2}}
\newcommand{\fracvalue}[2]{\fracdisp{#1}{\mathrm{#2}}}
\newcommand{\pma}[2]{{}^{+#1}_{-#2}}
\newcommand{\unit}[1]{\mbox{$\,\mathrm{#1}$}}
\newcommand{\nneib}{N_\mathrm{neigh}}
\newcommand{\msun}{M_{\sun}}
\newcommand{\mjup}{M_\mathrm{J}}
\newcommand{\tmsun}{\mbox{$\msun$}}
\newcommand{\tmjup}{\mbox{$\mjup$}}
\newcommand{\mjeans}{M_\mathrm{J}}

\newcommand{\csound}{c_\mathrm{s}}
\newcommand{\mdisk}{M_\mathrm{d}}
\newcommand{\mstar}{M_{\star}}
\newcommand{\mpert}{M_\mathrm{pert}}
\newcommand{\rperi}{r_\mathrm{peri}}
\newcommand{\mav}{\mbox{$M_\mathrm{\star}$}}
\newcommand{\mcl}{\mbox{$M_\mathrm{tot}$}}
\newcommand{\rpl}{\mbox{$r_\mathrm{Pl}$}}
\newcommand{\rhm}{\mbox{$r_{1/2}$}}
\newcommand{\rgv}{\mbox{$r_\mathrm{grav}$}}
\newcommand{\kms}{\mbox{$\mathrm{\,km\,s^{-1}}$}}
\newcommand{\tenc}{\mbox{$t_\mathrm{enc}$}}
\newcommand{\tcross}{\mbox{$t_\mathrm{cr}$}}
\newcommand{\nce}{\mbox{$n_\mathrm{enc}$}}
\newcommand{\psig}{q_{\Sigma}}
\newcommand{\pt}{q_{T}}
\newcommand{\ecc}{\mbox{$e$}}
\newcommand{\inc}{\mbox{$\iota$}}
\begin{document}
\title{Tidally induced brown dwarf and planet formation in circumstellar discs}
\author{Ingo Thies\altaffilmark{1}, Pavel Kroupa\altaffilmark{1},
Simon P. Goodwin\altaffilmark{2},
Dimitrios Stamatellos\altaffilmark3, Anthony P. Whitworth\altaffilmark3}
\altaffiltext{1}{Argelander-Institut f\"ur Astronomie (Sternwarte), Universit\"at Bonn, Auf dem H\"ugel 71, D-53121 Bonn, Germany}
\altaffiltext{2}{Department of Physics and Astronomy, University of Sheffield, Sheffield S3 7RH, UK}
\altaffiltext{3}{School of Physics \& Astronomy, Cardiff University, Cardiff CF24 3AA, UK}

\begin{abstract}
Most stars are born in clusters and the resulting gravitational
interactions between cluster members may significantly affect the
evolution of circumstellar discs and therefore the formation of
planets and brown dwarfs. Recent findings %\citep{FoRi09}
suggest that
tidal perturbations of typical circumstellar discs due to close
encounters may inhibit rather than trigger disc
fragmentation and so would seem to rule out planet formation by external
tidal stimuli.  However, the disc models in these calculations
were restricted to disc radii of 40~AU and disc masses below
0.1~\tmsun.  Here we show that even  modest encounters can trigger
fragmentation around 100~AU in the sorts
of massive ($\sim0.5\,\msun$), extended ($\ge100\unit{AU}$) discs
that are observed around young  stars.
Tidal perturbation alone can do this, no disc-disc collision is required.
We also show that very-low-mass binary systems can form through the
interaction of objects in the disc.  In our computations,
otherwise non-fragmenting massive discs, once perturbed, fragment
into several objects between about 0.01 and 0.1~\tmsun, i.e., over the
whole brown dwarf mass range. Typically these orbit on highly eccentric
orbits or are even ejected.  While probably not suitable for the
formation of Jupiter- or Neptune-type planets, our scenario
provides a possible formation mechanism for brown dwarfs and very 
massive planets which, interestingly, leads to a mass distribution
consistent with the canonical substellar IMF. As a minor outcome,
a possible explanation for the origin of misaligned extrasolar planetary
systems is discussed.
\end{abstract}
\keywords{%
binaries: general ---
open clusters and associations: general ---
stars: formation ---
stars: low-mass, brown dwarfs ---
planetary systems: protoplanetary disks ---
}
\maketitle

\section{Introduction}
The origin of brown dwarfs (BDs, to which we also include very
low-mass stars $<0.1$\tmsun) as well as the formation of massive
exoplanets is contentious. The most common idea is that BDs form 
like stars, and planets form by accretion onto a core formed from dust
conglomeration within a circumstellar disc.
However, there is increasing evidence that an  
alternative formation scenario for BDs
and, possibly, some of the most massive exoplanets is required. BDs 
differ from low-mass stars in several ways.
Firstly, there is the `brown dwarf desert' -- a lack of BD
companions to stars, especially at separations below
about 30~AU within which companion planets and stars
are common.  In addition, the properties of BD-BD binaries are
very different from star-star binaries, showing a significant lack of
systems wider than $\approx 20$~AU beyond which most stellar binaries
are found. Furthermore, while typically every second star is
a binary system only every fifth BD is a binary. Clearly
we must ask ourselves why stars and BDs are so different if they form
by the same mechanism. Indeed, in order to numerically set-up a realistic
stellar and BD population, stars and BDs need to be according to
different algorithmic pairing rules \citet{Ketal03}. In other words,
they follow different formation channels.

The fragmentation of extended circumstellar discs is one of the most
promising alternatives to the standard scenario of star-like formation
for BDs \citep{GoWi07}. Research on circumstellar disc evolution
has made great progress in recent years, but there are still many
unanswered questions \citep{PapTer06,Henning2008,Hillenbrand2008}.
One issue that has gained surprisingly little
attention in the past is the role of gravitational interactions
between the disc and passing stars within the young host stellar
cluster. Since stars are generally born in clusters \citep{FoRi09},
rather than as isolated objects, star-disc and disc-disc
interactions \citep{NSoPE1,NSoPE2,pfalzner:2005,TKT05} probably play an
important role for the dynamical evolution of such discs and thus for
massive planet and BD formation through fragmentation of the discs.

For a typical open star cluster hosting about
1000 stars within a half-mass radius of 0.5~pc, encounters closer than
500~AU will happen approximately every 10~Myr \citep{TKT05}.
They are even more frequent in denser clusters
in which are thought to form at least half of the stars in the Galaxy.
Encounters in star clusters are therefore expected to play a significant
role in the evolution of extended discs of several 100~AU radii. 

The formation of planets through core accretion is typically
expected to happen within the
innermost 100~AU of a circumstellar disc \citep{Hillenbrand2008},
although there are many unanswered questions especially
concerning the very first
stages of core accretion \citep{Henning2008}. Fragmentation has 
been proposed as an alternative way to form the most massive 
gas giant planets by \citet{Mayeretal:2002,Boss97,2004ApJ...610..456B,Boss06}.
In recent years, however, there has been a growing consensus that disc
fragmentation cannot produce gas giant planets in the inner regions of 
discs for two reasons: (1) heating from the central star stabilizes
the innermost regions and thus inhibits fragmentation,
and (2) these regions cannot cool fast enough to allow
temporary gas clumps to collapse \citep{Lodatoetal2007,StaWhi08,Caietal2010}.
This picture changes greatly beyond about
100~AU where cooling becomes sufficiently efficient for
gravitational collapse to occur. Consequently, disc fragmentation has been
proposed as an alternative mechanism for the formation of BDs
\citep{WhiSta06,GoWi07}. Most of these investigations, 
however, assume that the disc evolves in isolation.

Star-disc and disc-disc collisions have been investigated in previous
studies \citep{NSoPE1,NSoPE2,NSoPE3} but without a realistic treatment
of the radiative heat transfer in the disc. From SPH computations
\citet{FoRi09} deduce that close encounters inhibit fragmentation in
typical protoplanetary discs (assuming an initial radius of 40~AU) rather
than inducing it. In contrast to that, a recent study of disc-disc
collisions \citep{Shenetal10} has, however, shown that the direct
compression of gas by the close encounters of stars hosting massive
extended discs (disc mass $\approx 0.5$~\tmsun, disc radius $\approx
1000$~AU), may trigger the formation of substellar companions. However 
such massive, extended discs are probably short-lived \citep{EiCa06} 
(in contrast to typical protoplanetary discs which are about 
ten times smaller and less massive), so that pairwise interactions
of two discs of this kind are expected to be rare. The purely gravitational
interaction of a low-mass star with no
(or only a small) protoplanetary disc with a star hosting a massive disc
would be much more likely. Although even these encounters may probably
account for only a fraction of all BDs, especially in loose associations
like Taurus-Auriga, they provide an additional channel of forming BDs out
of discs that would otherwise probably dissolve without ever fragmenting.
In this paper we address this issue.
In Section~\ref{sec:model}, besides an estimate of the encounter probability,
the model of the circumstellar disc is described.
Section~\ref{sec:method} depicts the computational methods and gives an
overview of the model parameters studied in this work. The results are
then presented in Section~\ref{sec:results} and discussed in
Section~\ref{sec:discussion}

\section{Model basics}\label{sec:model}
\subsection{Encounter probability}
The likelihood of a close stellar encounter with an impact parameter $b$
depends on the stellar density of the star-forming environment and the
impact parameter itself.
The encounter
event is then further characterized by the relative velocity of the two stars
and thus the eccentricity of the path. The consequences on the circumstellar disc
further depend on the mass of the perturber and the host star (i.e., the
ratio of binding energy and magnitude of the perturbation) and the inclination
of the disc plane wrt. the plane of the encounter hyperbola.

Stars can be born in a variety of environments from low-density groups
like Taurus-Auriga (about 300 stars distributed in groups, each being <1 pc in
radius and containing about a few dozen stars) up to ONC-type clusters
(thousands of stars within 0.5~pc) or even in more massive and dense
clusters \citep{Kroupa05}.
An estimate of the encounter probability has been derived assuming a Plummer
star cluster model of half-mass radius, \rhm, total mass \mcl,
and an average stellar mass, \mav. Note that two other radial scales are
also often used to describe a Plummer model: the Plummer radius,
$\rpl=\sqrt{2^{2/3}-1}\rhm\approx0.766\rhm$,
and the gravitational scale radius,
$\rgv\approx2.602\rhm\approx3.395\rpl$.
From \citet{TKT05}, Section 1.3, the expected time, $\tenc$, between
two encounters within an encounter parameter, $b$, can be
obtained from the characteristic crossing
time \tcross,
\begin{equation}
\fracvalue{t_\mathrm{cr}}{Myr}=83\left(\frac{\mcl}{\msun}\right)^{-1/2}
\left(\fracvalue{\rhm}{pc}\right)^{3/2}\,,
\end{equation}
and the average number, \nce, of encounter parameters below $b$, for a given
number of stars, $N$,
\begin{equation}\nce(<b)=N\frac{b^2}{\rgv^2}\,.\end{equation}
$\tenc$ is approximately given by
\begin{equation}
\frac{\tenc}{\mathrm{Myr}}=\frac{2.4\dez{13}}{N}\left(\frac{\mcl}{\msun}\right)^{-1/2}
\left(\frac{\rhm}{\mathrm{pc}}\right)^{7/2}\left(\frac{b}{\mathrm{AU}}\right)^{-2}\,.
\end{equation}
It has to be noted that the actual periastron distance, $\rperi$, is always
(and sometimes significantly) smaller than $b$ due to gravitational focusing.
The relation between $\rperi$ and $b$ for a given eccentricity, $\ecc$, is
%$\rperi=\sqrt{(\ecc-1)/(\ecc+1)}\,b$.
\begin{equation}
\rperi=\sqrt{\frac{\ecc-1}{\ecc+1}}\,b\ .
\end{equation}
For an ONC-type cluster with $N=7500$, an average star mass $\mav=0.5\,\msun$,
a total mass $\mcl=10000\,\msun$ (i.e., 30\pct\ of the mass consists of
stars and the rest of gas, see \citealt{KAH01}), and a half-mass
radius of $\rhm=0.5\unit{pc}$
we find that encounters below 500~AU occur on average every 11 million years.
If a massive extended disc exists for about 1 Myr
this means that about 9 per cent of all discs of this type will suffer
such an encounter. See also \citet{TKT05}.

\subsection{Disc model}
\begin{figure}
\begin{center}
\plotone{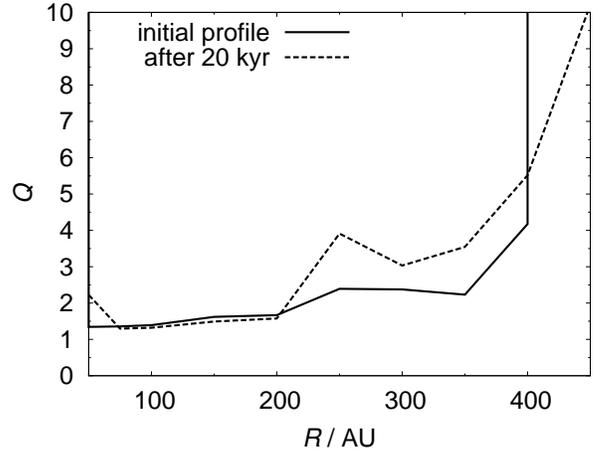}
\caption{\label{discprofile}The Toomre parameter $Q$ as a function
of the radius from the central star for both the initial disc setup
(solid line) and the settled disc after 20~kyr (dashed line).
The profiles of both disc stages are nearly identical below
about 200~AU while $Q$ is slightly larger (i.e. the disc is more stable)
between 200~AU and 400~AU, the outer rim of the initial disc.
The region within 50~AU is skipped since it is initially partially
void.}
\end{center}
\end{figure}

The initial conditions for the disc are taken from \citet{StaWhi08,StaWhi09a}.
The disc model has a power law profile for temperature, $T$, 
and surface density, $\Sigma$ from \citet{StaWhi08}, both as a
function of the distance, $R$, from the central star,
\begin{equation}\label{eq:sigma}
\Sigma(R)=\Sigma_0\,\left(\frac{R}{\mathrm{AU}}\right)^{-\psig}\,,
\end{equation}
\begin{equation}\label{eq:temp}
T(R)=\left[T_0^2\,\left(\frac{R}{\mathrm{AU}}\right)^{-2\pt}+T_\infty^2\right]^{1/2}\,,
\end{equation}
where $\psig=1.75$, $\pt=0.5$, and $T_0=300\unit{K}$.
$\Sigma_0$ is chosen corresponding to a disc mass of 0.48 and 0.50~\tmsun\
(see Table~\ref{tab:sims}), i.e., $\Sigma_0=8.67\dez{4}\unit{g\,cm^{-2}}$
and $\Sigma_0=9.03\dez{4}\unit{g\,cm^{-2}}$, respectively.
$T_\infty=10$~K is a background temperature to account for the background
radiation from other stars within the host cluster. 
This configuration leads
to an initial Toomre stability
parameter, $Q$, \citep{toomre:1964} between 1.3 and 1.4 between about
100 and 400~AU, leading to weak spiral density patterns but not to fragmentation.
$Q$ is given by
\begin{equation}
Q=\frac{\csound\kappa}{\pi G\Sigma}
\end{equation}
with $\csound$ being the sound speed, and $\kappa$ being the
epicyclic frequency which can be, at least roughly,
approximated by the Keplerian orbital frequency.
The radiation of the perturber is included in the same way as that
of the central star, assuming a mass-to-luminosity relation
of $L\propto M^4$ for low-mass stars. However, as long as the
perturber is no more than about 0.5~\tmsun, corresponding
to an early M-type or late K-type main sequence star, its
influence on disc evolution is very small.

The disc is initially populated by SPH particles between 40 and 400~AU.
Before starting the
actual encounter computation the disc is allowed to ``settle'' for about
20~000 years (i.e., about two orbits at the outer rim) to avoid
artefacts from the initial distribution function. During this settling the
disc smears out at the inner and the outer border extending its radial
range from a few AU up to $\gtrsim 500$~AU, eventually stabilising.
The $Q$ value is slightly increased beyond 200~AU relative to the
initial setting while remaining largely unchanged below this radius,
as can be seen in Figure \ref{discprofile}.
The actual surface density
after the settling is therefore slightly smaller and thus the disc slightly
more stable than in the initial setup. Any self-fragmentation of the
unperturbed disc would have happened during the settling time when the
Toomre parameter is lowest.

\section{Methods}\label{sec:method}

\subsection{Smoothed particle hydrodynamics (SPH)}
\label{ssec:sph}
All computations were performed by using the well-tested DRAGON SPH code
by \citet{Goetal04} including the radiative transfer
extension by \citet{Staetal07}. Most of the numerical
parameter settings have been adopted from \citet{StHuWi07,StaWhi09a}.
Gravitationally collapsing clumps are treated by sink particles which form
if the volume density exceeds $10^{-9}\unit{g\,cm^{-3}}$, and the clump
is bound. The sink radius
is 1~AU, in accordance with the studies mentioned above and
in agreement with the local Jeans criterion. In addition, the central
star and the perturbing star are both represented by sink particles.

There are ongoing discussions whether fragmentation can be triggered
artificially due to the numerical behaviour of SPH and grid-based
hydrocodes. \citet{HuGoWh06}
find that smoothed particle hydrodynamics (SPH) does not suffer
from artificial fragmentation. However, \citet{Nelson2006} 
suggest that SPH may artificially enhance fragmentation due to a
pressure-underestimation if the smoothing radius, $h$, is considerably
larger than about 1/4 of a vertical scale height of a circumstellar disc.
\citet{StaWhi09b} show that this criterion is fulfilled for
150,000 or more particles, for the types of disc studied in this paper.

\cite{CHACT08} have performed
calculations of fragmenting protostellar clouds with the adaptive
mesh refinement (AMR) code RAMSES and the SPH code DRAGON (which
we have been using for this study) and compared the results at 
different resolutions given by the number $N$ of grid cells and particles,
respectively. For about $N=5\dez{5}$ cells/particles
fragmentation in both codes appeared nearly equal.
With $N=2\dez{5}$ the results were very similar to the
high-resolution AMR computations while being still acceptable
for many purposes at $N=1\dez{5}$. \citet{StaWhi09b} briefly review
the resolution criteria with respect to their models using 150~000 in that
paper and 250~000 to 400~000 particles earlier \citep{StaWhi09a}.

The three most important resolution criteria are the resolution of
the local Jeans mass, $\mjeans$, (and thus the Jeans length), the local
Toomre mass and the vertical scale
height of the disc. In particular, the local Jeans mass,
\begin{equation}
\mjeans=\frac{4\pi^{5/2}}{24}\,\frac{\csound^3}{\sqrt{G^3\,\rho}}\,,
\end{equation}
must be resolved by at least a factor of $2\times\nneib=100$,
corresponding to 0.5~\tmjup\ for the 100,000 particle discs
and 0.2~\tmjup\ for the 250,000 particle discs.
Since the global disc setup as well as the evolution after the
perturbation (Section \ref{sec:results}) is quite similar to that used by
\citep{StaWhi09b}, as similar range of Jeans masses of the forming
clumps, i.e., $\sim 2$--$20$~\tmjup, can be reasonably assumed
for our models and has actually been tested for
the models E/X002 and E/X009 for all forming objects.
The minimum $\mjeans$ during the clump formation is
typically between 4 and 6~\tmjup\ and does never go below about
2~\tmjup.
Therefore, $\mjeans$ is resolved by a safe factor of at least 4
in the low-resolution case and 10 in the high-resolution case.
Accordingly, the minimum Toomre mass of 2.5~\tmjup\ is adequately
resolved as well as the vertical scale height (by at
least a few smoothing lengths).

\subsection{Radiative transfer model}
In \citet{TKT05} the possibility of
fragmentation has been estimated via the Toomre criterion for both
isothermal and adiabatic equations of state, both resulting in
perturbed regions with highly unstable conditions ($Q<1$).
However, Toomre instability does not necessarily lead to actual
fragmentation.

In realistic models of radiative transfer, such as the one used here,
the thermal response of the gas to density changes is between near-isothermal
in regions with efficient cooling (outside about 100--200~AU)
and near-adiabatic in regions with long cooling times (the inner parts
of the disc). \citet{Shenetal10} have shown that
direct disc-disc collisions may lead to fragmentation of massive
extended discs at some 400~AU even for a ``thick disc approximation'',
i.e., the near-adiabatic case.

The \citet{Staetal07} method uses the density and the 
gravitational potential of each SPH particle to estimate an optical depth 
 for each particle through which the particle cools and heats. The 
method takes into account compressional heating, viscous heating, 
radiative heating by the background, and radiative cooling. It performs 
well, in both the optically thin and optically thick regimes, and has been 
extensively tested by \citet{Staetal07}. In particular it reproduces 
the detailed 3D results of \citet{MasInu2000}, \citet{BossBodenheimer79},
\citet{BossMyhill92}, \citep{WhitehouseBate06}, and 
also the analytic results of \citet{Spiegel57}. Additionally the code has been 
tested and performs well in disc configurations as it reproduces the 
analytic results of \citet{Hubeny90}.

The gas is assumed to be a mixture of hydrogen and helium. We use an equation
of stare by \citet{BlackBodenheimer75}; \citet{Masunagaetal98}; \citet{Boleyetal07}
that  accounts (1) for the rotational and vibrational degrees of freedom of 
molecular hydrogen, and (2) for the different chemical states of hydrogen 
and helium. We assume that ortho- and para-hydrogen are in equilibrium.

For the dust and gas opacity we use the parameterisation  by \citet{BellLin94},
$\kappa(\rho,T)=\kappa_0\ \rho^a\ T^b\,$, where $\kappa_0$, $a$, 
$b$ are constants that depend on the species and the physical processes 
contributing to the opacity at each $\rho$ and $T$. The opacity changes 
due to ice mantle melting, the sublimation of dust, molecular and H$^-$ 
contributions,  are all taken into account.

\subsection{Overview of Calculations}
\begin{table*}\resizedraft
\begin{center}
\tabhead{\label{tab:sims}Calculations Summary}
\begin{tabular}{c cccc cccc ccc}\tableline%\mbox{()}
Model &$N/1000$&$\tabunit{\mstar}{\msun}$&$\tabunit{\mdisk}{\msun}$&$\tabunit{\mpert}{\msun}$&$\tabunit{\rperi}{\mbox{AU}}$&$\ecc$&$\tabunit{\inc}{\mbox{deg}}$&$\tabunit{m_\text{env pert}}{\mjup}$&$N_\text{formed}$&$N_\text{ejected}$&Binaries?\\\tableline
E002  &100     &1.00          &0.50&0.50& 500 & 1.1 & 10 &11.7&4&1&---\\
X002  &250     &1.00          &0.50&0.50& 500 & 1.1 & 10 &12.0&5&2&1 ($\approx 4$~AU)\\
Y002  &250     &1.00          &0.50&0.50& 500 & 1.1 & 10 &10.2&3&---&---\\
E003  &100     &0.75          &0.48&0.50& 500 & 1.1 & 10 &16.1&3&2&1 ($\approx 2$~AU)\\
X003  &250     &0.75          &0.48&0.50& 500 & 1.1 & 10 &12.6&4&1&---\\
E004  &100     &0.75          &0.48&0.50& 500 & 1.5 & 10 & 5.6&2&---&---\\
X004  &100     &0.75          &0.48&0.50& 500 & 1.5 & 10 &11.0&3&---&---\\
E006  &100     &0.75          &0.48&0.50& 500 & 2.0 & 10 & 4.8&2&---&---\\
X006  &250     &0.75          &0.48&0.50& 500 & 2.0 & 10 & 4.4&3&---&1 ($\approx 4$~AU)\\
E007  &100     &0.75          &0.48&0.50& 500 & 3.0 & 10 & 2.1&0&---&---\\%\tableline
E008  &100     &0.75          &0.48&0.50& 600 & 1.5 & 10 & 2.8&0&---&---\\%\tableline
F008  &100     &0.75          &0.48&0.50& 600 & 1.1 & 10 & 6.5&0&---&---\\%\tableline
E009  &100     &0.75          &0.48&0.50& 500 & 1.5 & 30 & 3.0&3&---&---\\
X009  &250     &0.75          &0.48&0.50& 500 & 1.5 & 30 & 7.6&4&---&1\\
E010  &100     &0.75          &0.48&0.50& 550 & 1.5 & 10 & 4.1&3&---&---\\
X010  &250     &0.75          &0.48&0.50& 550 & 1.5 & 10 & 8.8&4&---&---\\%\tableline
E011  &100     &0.75          &0.48&0.50& 500 & 1.5 & 45 & 7.6&6&---&---\\
F011  &100     &0.75          &0.48&0.50& 500 & 1.5 & 45 & 4.2&2&---&---\\
X011  &250     &0.75          &0.48&0.50& 500 & 1.5 & 45 & 6.1&2&---&---\\%\tableline
E015  &100     &0.75          &0.48&0.40& 400 & 1.5 & 45 & 7.9&3&---&---\\
E016  &100     &0.75          &0.48&0.30& 400 & 1.5 & 30 &10.4&2&---&---\\\tableline
\multicolumn{9}{r}{Total $N=100\,000$ (10 of 13 models showing fragmentation):}&30&3 &1  \\%\tableline
\multicolumn{9}{r}{Total $N=250\,000$ (8 of 8 modes showing fragmentation):} &28&3 &3  \\%\tableline
\multicolumn{9}{r}{Grand total (18 of 21 models showing fragmentation):}      &58&6 &4  \\\tableline
\end{tabular}\\[1ex]
\tablecomments{%
Overview of our SPH computations. The columns (from left to right)
show the computation ID, the number of gas particles in thousands,
the mass of the central star, its disc and the perturber star. Then follows
the periastron distance of the encounter, the eccentricity and inclination.
The four rightmost columns show the amount of gas captured by the perturber
in a circumstellar envelope within a radius 40~AU,
the number of formed companions in total and the number of bodies that
got ejected during the calculation as well as the number of binary systems.
In all models except for E015 and E016
the perturbing star has 0.5~\tmsun\ and passes the central star on
an initially hyperbolic orbit with initial eccentricity between 1.1 and 3.0
and inclination
to the disc plane between 10\textdegree\ and 45\textdegree.}
\end{center}
\end{table*}

We have conducted thirteen SPH calculations 
using 100,000 and eight using 250,000 SPH particles (i.e., 21 models in total)
corresponding to about nine CPU months in total on 16-core machines.
The model identifiers that begin with an ``E'' or ``F'' refer to the
low resolution models while the models with ``X'' or ``Y'' are the
high resolution ones. Subsequent letters with identical numbers
correspond to follow-up calculations (beginning at the moment of
the closest encounter) with identical model settings. Due to the dynamical
interaction of the parallel computing CPUs a slight random perturbation
is imposed to the subsequent evolution such that the outcome differs within
the statistical noise.
The high resolution calculations are set up with parameters of preceding
lower-resolution models that showed fragmentation. Therefore, the fact that
all high resolution computations show fragmentation is due to the parameter
selection. 
The low-resolution calculations are used as a parameter
survey while the high resolution ones are follow-ups to selected low-resolution
ones. Please note that not all non-fragmenting models are listed
here but only those which represent the borders of the parameter space
between fragmentation and no fragmentation.

We chose a 0.5~\tmsun\ perturbing star as a typical member of a star
cluster \citep{Kr01} for the majority of models, but we also performed
calculations with perturbers down to 0.3~\tmsun.
The encounter orbit is slightly inclined
($\inc=10\grad$ against the initial disc plane) with varying
eccentricities from $\ecc=1.1$ (near-parabolic) to $\ecc=2$ (hyperbolic),
corresponding to a relative velocity of 0.4--2.7\kms\ at infinity.
For these parameters the disc typically fragments to form a few very-low-mass
stars and substellar-mass objects.
A calculation with $\ecc=3$ has been performed but yielded no fragmentation.
The same holds true for encounters at 600~AU or more.
Depending on the eccentricity the calculations start 5000--10000 years before
periastron to ensure a sufficiently low initial interaction.
The calculations
continue for 15~000 years after encounter for the models E/X/Y002 and
at least 20~000 for the others. The masses and orbital radii of the objects
formed in the calculation are determined at this time.
After 15~000 years the accretion process of the objects
has largely finished while dynamical evolution may still lead
to major changes of the orbital parameters (and even to ejections of some bodies)
if the calculations would have been continued over a longer time interval.
For this reason, the orbital separations computed in this
study have to be taken as a preliminary state.

It has to be noted that the orbit is altered slightly during the passage
due to the transfer of mass and angular momentum.
Furthermore, since the encounter dissipates energy
the post-encounter speed is typically lower than the 
pre-encounter speed and in some
cases both stars may be captured in an eccentric binary (which
actually happened to the companions 1a and 1b in calculation E003;
see Table~\ref{tab:sims} and Figures \ref{imgxx02} and \ref{imge003}).
We found, however, that these effects are small.

\section{Results}\label{sec:results}
%%%%%%%%%%%%%%%%%%%%%%%%%%%%%%%%%%%%%%%%%%%%%%%%%%%%%%%%%%%%%%%%%%%%%%%%%%%%%%%%
\begin{table}\resizedraft
\begin{center}
\tabhead{\label{tab:results}Outcome of Individual Calculations}
\begin{tabular}{ccc@{\quad}ccc}
\multicolumn{3}{c}{100~000 particles}&\multicolumn{3}{c}{250~000 particles}\\\tableline
\tabover{\mbox{Model~ID}}{\mbox{E002}}&$\D\tabunit{m}{\msun}$&$\D \tabunit{\mbox{Separation}}{\mbox{AU}}$%
&\tabover{\mbox{Model~ID}}{\mbox{X002}}&$\D\tabunit{m}{\msun}$&$\D \tabunit{\mbox{Sep.}}{\mbox{AU}}$\\\tableline
1   &0.021    &332--342       &1   &0.013    &ejected       \\
2   &0.080    &ejected        &2   &0.046    &ejected       \\
3   &0.100    &20--30         &3   &0.057    &155--225      \\
4   &0.131    &50--100        &4a  &0.088    &70--110       \\
&&                            &4b  &0.099    &70--110       \\\tableline  
E003&\multicolumn{2}{c}{}     &X003&\multicolumn{2}{c}{}    \\\tableline
1a  &0.013    &130--500       &1   &0.021    &50--100       \\
1b  &0.086    &130--500       &2   &0.022    &200           \\
2   &0.130    &29--42         &3   &0.055    &ejected       \\
&&                            &4   &0.154    &130--200      \\\tableline 
E004&\multicolumn{2}{c}{}     &X004&\multicolumn{2}{c}{}    \\\tableline
1   &0.134    &160--190       &1   &0.026    &150--220      \\
2   &0.162    &60--90         &2   &0.096    &50--170       \\
    &&                        &3   &1.032    &25            \\\tableline
E006&\multicolumn{2}{c}{}     &X006&\multicolumn{2}{c}{}    \\\tableline
1   &0.063&35--70             &1a  &0.033    &150           \\
2   &0.154&120--220           &1b  &0.046    &150           \\
&&                            &2   &0.148    &60            \\\tableline
E009&\multicolumn{2}{c}{}     &X009&\multicolumn{2}{c}{}    \\\tableline
1   &0.026&16                 &1   &0.049    &20           \\
2   &0.099&50                 &2   &0.064    &230          \\
3   &0.133&140                &3   &0.073    &230          \\
&&                            &4   &0.138    &60           \\\tableline
E010&\multicolumn{2}{c}{}     &X010&\multicolumn{2}{c}{}   \\\tableline
1   &0.066&190                &1   &0.031    &190          \\
2   &0.117&80                 &2   &0.058    &40           \\
2   &0.092&80                 &2   &0.106    &40           \\
&&                            &3   &0.128    &80           \\\tableline
E011&\multicolumn{2}{c}{}     &X011&\multicolumn{2}{c}{}   \\\tableline
1   &0.031&110                &1   &0.073    &100--180\\
2   &0.038&300                &2   &0.107    &50--80  \\
3   &0.038&700                &&&\\
4   &0.048&230                &&&\\
5   &0.074&15--20             &&&\\
6   &0.098&125                &&&\\\tableline
F011&\multicolumn{2}{c}{}     &Y002&\multicolumn{2}{c}{}     \\\tableline
2   &0.090&150--400           &1   &0.080    &$\approx110$   \\
2   &0.168&60--100            &2   &0.108    &100--600       \\
&&                            &3   &0.137    &$\approx50$    \\\tableline
\multicolumn{6}{c}{Additional calculations with 100~000 particles}\\
E015&\multicolumn{2}{c}{}     &E016&\multicolumn{2}{c}{}     \\\tableline
1 & 0.070 & ejected           &1 & 0.093 & 13--30            \\
2 & 0.103 & 10--29            &2 & 0.149 & 55--98            \\
3 & 0.134 & 70--101           &&&\\\tableline

\end{tabular}\\[1ex]
\tablecomments{%\sffamily\small
List of the companions formed during individual calculations
within 15~000--20~000 years after the flyby, sorted by mass. The table shows
the masses and the approximate minimum and maximum separations from the
central star.
Note that bodies 4a and 4b in X002 got bound in a VLMS binary with a
mutual distance of about 4~AU through a triple encounter
while 1a and 1b in E003 got bound through dissipational
capture.
The binary capture of 1a and 1b in X006 probably involved both mechanisms.}
\end{center}
\end{table}

%%%%%%%%%%%%%%%%%%%%%%%%%%%%%%%%%%%%%%%%%%%%%%%%%%%%%%%%%%%%%%%%%%%%%%%%%%%%%%%%
\begin{figure}
\begin{center}
\plotone{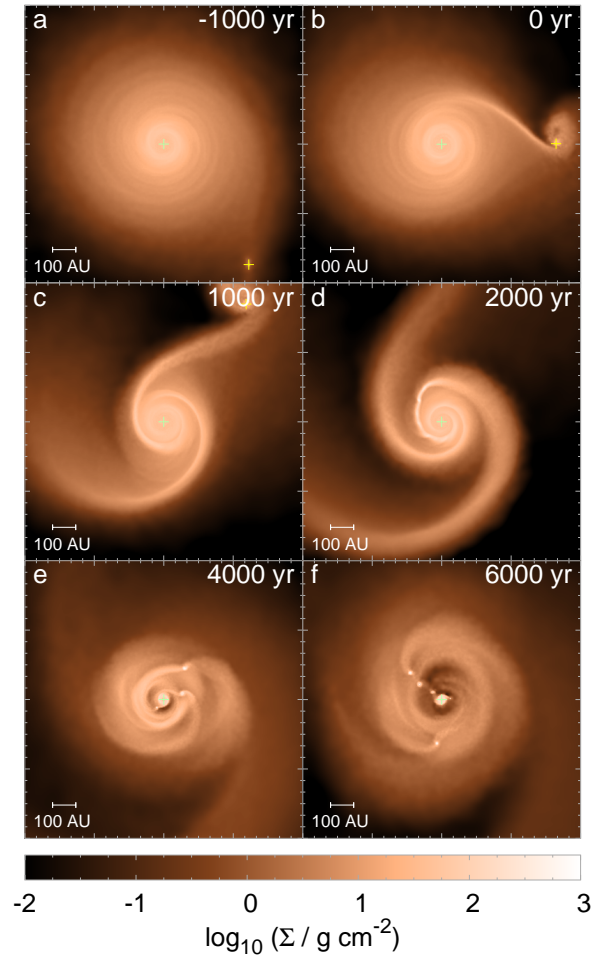}
\caption{\label{imgx002}
Snapshot of a circumstellar disc modelled with 250,000 SPH particles,
around a Sun-type star being perturbed by a close star-star encounter
(model X002, Table~\ref{tab:sims}).
The time stamp in each frame
refers to the
time of the encounter.}
\end{center}
\end{figure}

\begin{figure}
\begin{center}
\plotone{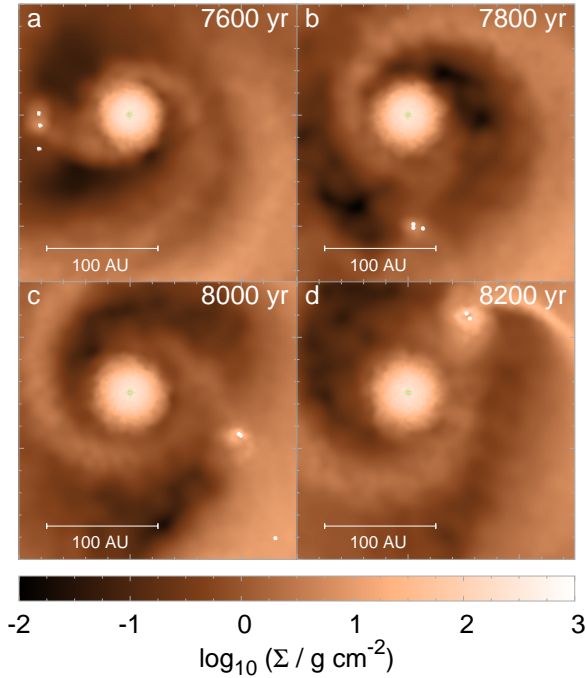}
\caption{\label{imgxx02}
Snapshots of a forming binary around about 8000 years
after the fly-by in model X002 (see also Figure \ref{imgx002}).
The components of the remaining VLMS pair have masses of 0.08~\tmsun\ and
0.09~\tmsun, subsequently accreting another 0.01~\tmsun\ each. The escaping
third body has 0.05~\tmsun\ and is eventually ejected.}
\end{center}
\end{figure}

\begin{figure}
\begin{center}
\plotone{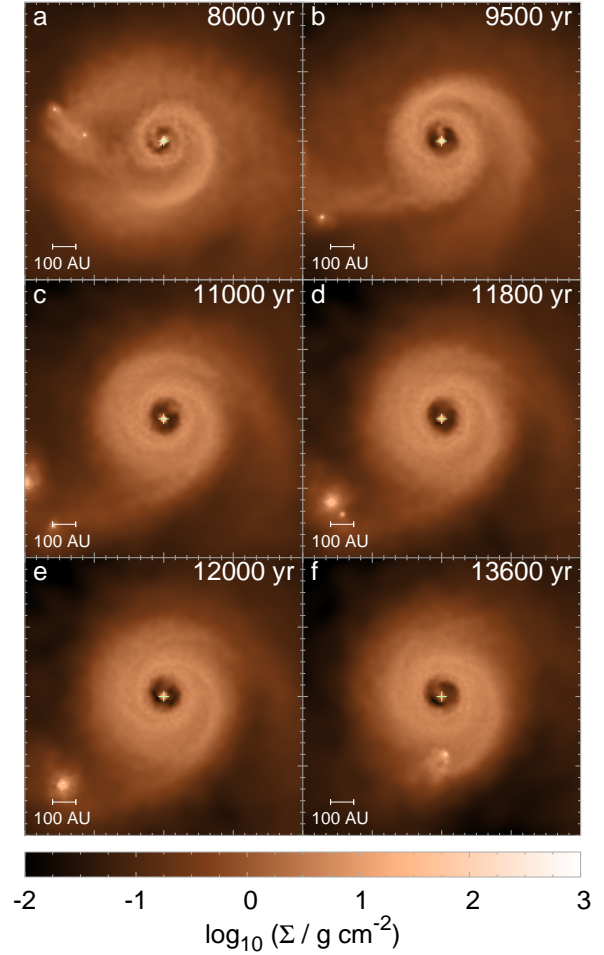}
\caption{\label{imge003}
Snapshots of another forming binary around about 12000 years
after the fly-by.
The components of the remaining VLMS-BD pair have very unequal
masses of 0.09~\tmsun\ and 0.013~\tmsun\ (with a mass ratio $q=0.14$).
In contrast to the binary formed in X002 (Figure \ref{imgx002})
these bodies became bound due to a frictional
encounter of their accretion envelopes.}
\end{center}
\end{figure}

\subsection{General findings}
\begin{figure}
\begin{center}
\plotone{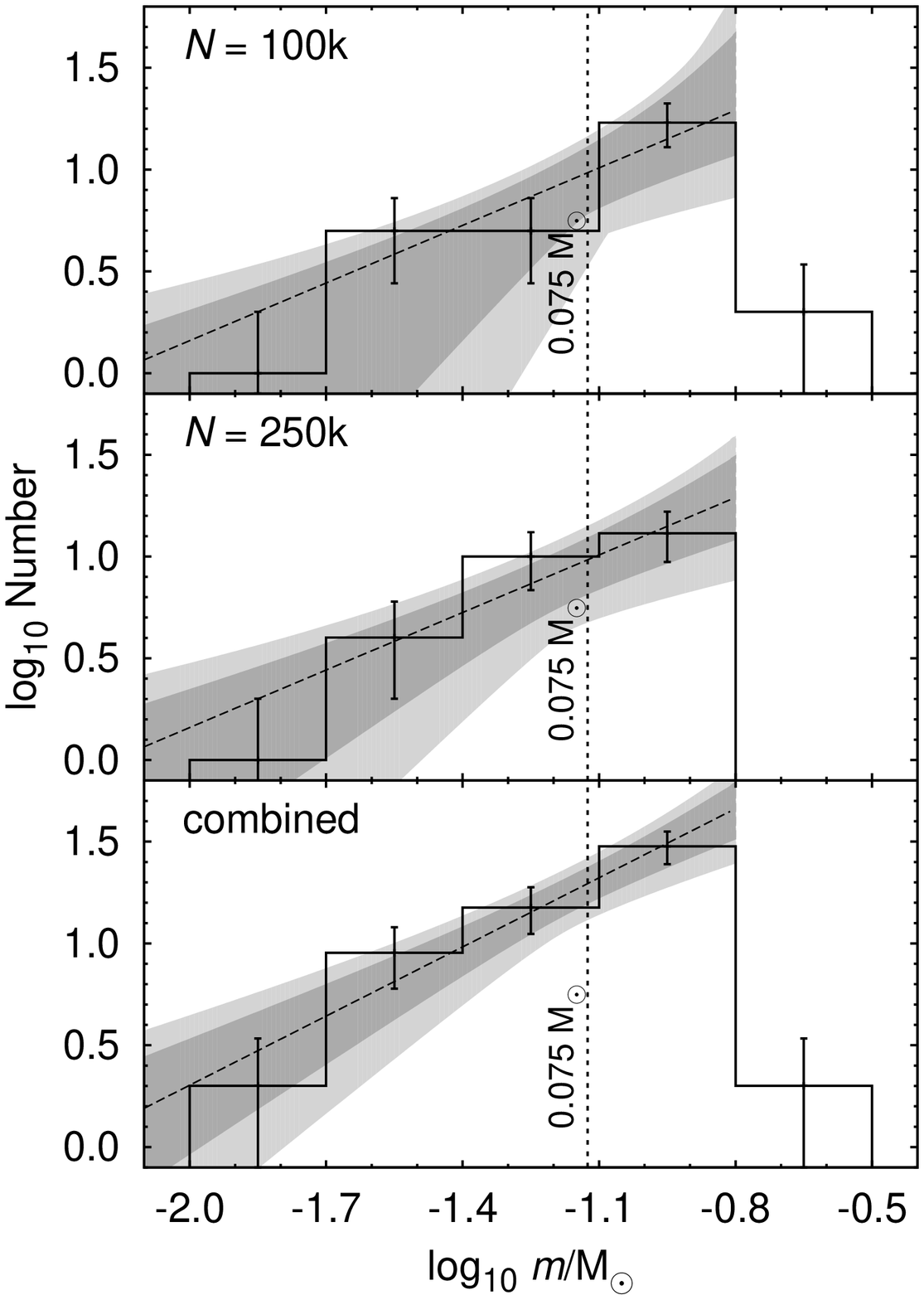}
\caption{\label{created}
Mass distribution of the bodies created in the 100~000 particle
models (top frame), the 250~000 particle models (middle frame),
and both combined (bottom frame).
The majority of companions is found in the mass interval
between 0.08 and 0.16~\tmsun\ while most of the rest is in the substellar
regime. The substellar mass function corresponds to a power law
index of $\alpha=+0.1\pma{0.3}{1.5}$ for the 100k models,
$\alpha=+0.1\pma{0.3}{0.6}$ for the 250k models, and
$\alpha=-0.1\pma{0.3}{0.4}$ for both combined (see text for further
explanations). The dark and light grey-shaded regions refer to the
1 and 2~$\sigma$ confidence limits of the fit while the errorbars
correspond to the 1~$\sigma$ Poisson errors.
The vertical dotted line marks the hydrogen-burning mass
limit of 0.075~\tmsun\ \citep{ChaBa00}.}
\end{center}
\end{figure}
%%%%%%%%%%%%%%%%%%%%%%%%%%%%%%%%%%%%%%%%%%%%%%%%%%%%%%%%%%%%%%%%%%%%%%%%%%%%%%%%

As the unperturbed disc is marginally stable it does not
fragment until the passage of the perturbing star. The first
visible effects of a typical fly-by are the appearance of tidal arms
and a mass transfer to the
passing star (which acquires a small disc). About 2000--3000
years after the periastron passage parts of the tidal arms become
gravitationally unstable, and spiral-shaped over-densities begin
to form (see the snapshots from model X002 in
Figure \ref{imgx002}\footnote[1]{Supplementary content like movies
from our calculations can be downloaded from the AIfA download page,
\mbox{http://www.astro.uni-bonn.de/\symbol{126}webaiub/german/downloads.php}
}).
Some of these continue to contract into a runaway collapse and form
bound objects represented by sinks (see Section~\ref{sec:method}).
This typically happens at radii between 100 and 150~AU from the
central star with some clumps forming even around 200~AU. Temporary
overdensities do also occur within less than 100~AU but, however,
dissolve quickly. This is probably due to the heating from the
central star and the less effective cooling in these regions as already
being mentioned in the Introduction and by \citet{WhiSta06,GoWi07}.
At radii larger than about 200~AU, on the other hand, the gas density
is apparently too low for gravitationally bound clumps for form.

Each of the calculations typically produces between two and five
(six on one case)
low-mass objects with masses between those of very massive planets
(0.01~\tmsun) and very low-mass
stars (0.13~\tmsun). Additionally, mass is accreted onto the central star,
and also mass is transfered to the perturber during the encounter.
In the model shown in Figure \ref{imgx002}
five objects with masses between 0.013~\tmsun\ and 0.10~\tmsun\ form;
two of them being bound in a binary system. In addition,
0.016~\tmsun\ of gas accretes onto the central 
star within 15~000 years, and 0.0028~\tmsun~=~2.8 Jupiter masses (\tmjup)
of gas transfers to the perturber (into the representing sink particle)
while about 0.012~\tmsun\ ($\approx12\,\mjup$) is captured
around the perturber as a circumstellar envelope. In our analysis,
any gas that is present within a radius of 40~AU but outside the sink
radius is counted as circumstellar material of the perturber and
shown in the ninth column of Table ~\ref{tab:sims}.
This mass capture is typical, ranging
between 2 and 16~\tmjup.

In total, 28 bodies between 0.01 and 0.15~\tmsun\ formed
in eight high resolution calculations,
while the ten 100~000 particle calculations that showed fragmentation,
yielded 30 bodies between 0.01 and 0.16~\tmsun\ (see Tables~\ref{tab:sims}
and \ref{tab:results}). Thus, there is no statistically significant relation
between the resolution and the number of formed bodies.
Similarly, no clear trend towards lower minimum masses of the companions
for the high resolution can be derived from the current data.
The average mass of the lowest-mass member is 0.056~\tmsun\ for the 100~000 particle
models while it is 0.044~\tmsun\ for the 250~000 ones. Similarly, the average
mass of the highest-mass is 0.097~\tmsun\ and 0.083~\tmsun, respectively.
Table \ref{tab:results} does also show the separations of the objects at the end
of each calculation. It has to be noted that these can differ largely from the
initial separation at the moment of formation. Dynamical interaction between the
objects and the disc as well as mutual encounters push some bodies at wide and
eccentric orbits or even eject them while others migrate closer to the star.
The closest separation observed in our calculations is about 10~AU in model
E015 after 15~000 years of evolution. The overall outcome is very similar
to that of unperturbed self-fragmenting disc as shown in \citet{StaWhi09b}
except for the lower disc mass. This is quite plausible since the main difference
between the two scenarios is the cause for the density patterns which undergo
fragmentation. While Toomre density waves are the source of fragmentation
in self-fragmenting discs, the gravitationally induced fragmentation occurs in
tidal arms. In both cases, a dense bar-like in the disc reaches the density
limit for fragmentation and thus the underlying physics is essentially
the same.

Figure \ref{created} shows the mass distribution of the
$n$ bodies created all calculations with 100~000 particles
(top frame), 250~000 particles (middle frame) and both combined
(bottom frame). A power law mass function
$dn/dm=k(m/\msun)^{-\alpha}$ can be fitted to this distribution. This has
been done in this figure for the substellar regime (dashed line). In the
bi-logarithmic scaling, the slope of $d\log n/d\log m$ corresponds
to $1-\alpha$ due to the differentiation of the logarithm
(see \citealt{TK07} for details).
The linear fit to the mass function for all calculation outputs combined
appears to have a slope of
$0.9\pma{0.3}{0.4}$ ($<0.16$~\tmsun)
corresponding to power law index of
$\alpha=-0.1\pma{0.3}{0.4}$.
If the calculations are separated by resolution,
the 100~000 particle models correspond to $\alpha=+0.1\pma{0.3}{1.5}$ while
the 250~000 particle models similarly yield $\alpha=+0.1\pma{0.3}{0.6}$.
The slightly flatter IMF in the separate mass functions may be interpreted
as a weak resolution-dependence of the average created mass which gets
smeared out in the combined mass function. However, the difference
between the slopes of $0.2$ is within the $1\sigma$ uncertainty
The uncertainty,
based on the Poisson errors of the log mass bins, is indicated by a shaded region.
Above 0.12~\tmsun\ there is a sharp drop in the mass distribution that
can be treated as a truncation. Similar results have been obtained by
\citet{StaWhi09a} and \citet{Shenetal10}.
Interestingly, this is also in good agreement with both the sub-stellar IMF
deduced by \citet{Kr01} as well
as with the separate substellar IMF deduced in \citet{TK07,TK08}.
However, more calculations are needed to provide statistically robust
tests of the IMF. Furthermore, there is no weighting with respect to the
different likelihood of the different encounter settings (i.e., inclination,
fly-by distance etc.). Since the outcome of the computations does not show a
specific dependency on the orbital parameters (except for whether there is
fragmentation) such a weighting might introduce an artificial bias by
amplifying the random noise of higher-weighted models.

\subsection{Binary formation}\label{ssec:binfrac}
In the same way as in the models of \citet{StaWhi09a} of isolated
circumstellar discs, very low-mass binary systems can also form
in gentle three-body encounters between low-mass objects in the disc.  
Such a triple in-orbit encounter occurred during the
X002 computation about 8,000 years after the fly-by, as shown in
Figure \ref{imgxx02} where a 4~AU binary system composed of 0.09~\tmsun\ and
0.10~\tmsun\ objects forms through a triple encounter.  This binary 
remains bound to the host star in an orbit of about 100~AU.
These separations are consistent with the observations of VLMS binaries
according to which the most probable separation is around 3~AU
(see, e.g., Figure 10 in \citealt{StaWhi09a}).
Another binary-forming mechanism was unveiled during calculation E003
(see Table~\ref{tab:results}). A 2~AU binary, consisting of a VLMS of
0.09~\tmsun\ and a BD with 0.013~\tmsun, formed through a grazing
encounter of circum-substellar accretion envelopes, which subsequently
evolve into a circumbinary disc. The event is shown in a sequence plot
in Figure~\ref{imge003}. In two similar events in run X006, two accreting
clumps merged to a single one of about 0.03~\tmsun\ 9600~yrs after the
fly-by, and 6000~yrs later this body and a third one got captured into
a binary of about 5~AU separation, probably involving both triple
encounter and grazing collision. The masses at the end of the calculation
are 0.03 and 0.04~\tmsun.

Binaries formed via these processes may later be separated from their host star
in a subsequent encounter with another star as already discussed for single
substellar companions \citep{Kr95a,GoWi07}, or, possibly, by dynamical
interaction with more massive companions in the same system.
These models produce 55 systems of BDs and VLMSs, and 3 binaries giving a
binary of $3/55=0.05$, whereby the binaries have semi-major axes of
4, 2 and 5~AU. Thus, while the separations are quite consistent with
the observed VLMS and BD binaries, the present theoretical binary fraction
is somewhat low.

\subsection{The role of eccentricity and inclination}
In a previous study \citep{TKT05} we found that the strength of the
perturbation increases with decreasing eccentricity and inclination.
This agrees with results of an SPH parameter study by \citet{pfalzner:2005}.
While disc-disc collisions produce most objects due to shock
formation \citep{Shenetal10}, coplanar encounters increase the
effectiveness of the tidal perturbations which cause our discs to
fragment. This is a
consequence of the low relative velocity between the perturber and the
perturber-facing parts of the disc in coplanar encounters, 
and thus a longer tidal exposure
time. While drag forces in disc-disc collisions are larger for larger collisional
velocities the tidal force does not depend on the velocity, and therefore the
total amount of tidally transferred momentum for a given flyby distance
is larger for slower encounters, i.e., for those with lower eccentricity.

\section{Discussion}\label{sec:discussion}
Our computations show that the tidal perturbations of massive 
(otherwise stable) discs can form massive planets and BDs.  
Previous computations had assumed the interaction of two massive 
discs \citep{Shenetal10}, reducing the probability of such an event
dramatically. Our scenario only requires a single extended disc.
This scenario can produce very low-mass companions at large distances
from the primary star, and may help explain recent direct detections of
massive planets orbiting at $>100$~AU, far beyond where core 
accretion could have formed them\footnote[2]{See also the Extrasolar Planets Encyclopaedia,
http://exoplanet.eu/}, as well as intermediate \citep{Liuetal02} and
distant \citep{StHuWi07,StaWhi09a} BD companions
to stars.  Like \citet{StaWhi09a} we are also able
to form BD binaries through three body encounters within the disc, and
in addition, through dissipative encounters. But the required disc mass
is considerably smaller, by about 30--40\pct, since even initially stable
discs do fragment upon perturbation in our calculations. The mass function
of companions formed by this process is in good agreement with that of
self-fragmentation and with the separate substellar IMF deduced in
\citet{TK07}. However, the volume of the current results does not
allow robust statistical tests of the normalization of this sub-stellar IMF
relative to the stellar IMF. Nevertheless, the obtained results are in
remarkable agreement both with the form of the substellar IMF
and the binary properties in the BD and VLMS regime, although the
binary fraction is somewhat low (Section \ref{ssec:binfrac}).
More computations will have to be
performed at different resolutions (up to a million particles) to enhance
quantity as well as quality of the data.

\subsection{How often do such encounters occur?}
Furthermore, it has to be noted that triggered fragmentation
is probably responsible only for a fraction of disc-fragmentation events
while a good fraction may be triggered simply by over-feeding of an accreting
disc towards self-fragmentation. Although the borders of the parameter
subspace suitable for triggered fragmentation are not yet fully
determined the current findings show that encounters of a higher inclination
than 45 degrees, outside 600~AU or with eccentricities above 2
are generally unlikely to trigger fragmentation for the disc type
studied here. The same holds true for perturbers
of less than 0.3~\tmsun.

By combining all these limits of the parameter space one can estimate the
fraction of random encounters with a mutual periastron below 500~AU that
are suitable for fragmentation of the analysed disc type. If a characteristic
velocity dispersion of $\sim2\unit{km\,s^{-1}}$ and an upper stellar mass limit
of 10~\tmsun\ within the host cluster is assumed
only about 3~\pct\ of all encounters below 500~AU lead to fragmentation.
However, discs with even only slightly larger mass or lower background
temperature may be pushed much easily to fragmentation, probably enlarging
the parameter space significantly. Furthermore,
other perturbing effects like stellar winds or supernova shock waves,
which might influence large volumes within the host cluster at once, are
not covered by this study, nor are the effects of dust.

\subsection{Consequences for planet formation}
All objects formed in our calculations have masses above 0.01~\tmsun\ or
10~\tmjup\ and are thus above the masses typically assumed for planets. However,
we cannot rule out at the moment that also objects below 0.01~\tmsun\
may form through fragmentation, especially around lower-mass host stars
which heat the inner disc region much less than more massive ones.
Another critical point may be the resolution limits (although the local
Jeans mass is well resolved even in the inner disc region, as discussed in
Section \ref{ssec:sph}).

It has to be emphasized, that planet formation probably
typically takes place in the inner disc region through core-accretion
\citep{Hillenbrand2008} which are less influenced by the perturbation
unless companions formed through it migrate into these inner regions.
There might still be significant impacts on the outcome of planet formation
for stellar encounters, though.
Even temporary gravitational instabilities that do not collapse into a
brown dwarf or planet directly may induce the formation of substellar companions
down to Kuiper-Belt objects by induced vorticity and subsequent dust
trapping \citep{BaSo1995,KlahrBodenheimer2003,KlahrBodenheimer2006}.
Also the development of baroclinic vortices may be altered under the influence
of tidal perturbations, either inhibiting or promoting the formation of
dust aggregates.
This mechanism may even work in typical protoplanetary discs and is subject
of our ongoing research. According to solar system architecture
\citep{heller:1993,eggersetal:1997,2004Natur.432..598K} and
radionuclide evidence (\citealt{Takigawaetal2008,SaGu09,Gaidosetal2009};
\citealt{GuoMei08}, however, disagree)
the highly probable origin of our Sun in a large
star-forming region  further emphasizes the importance of such scenarios.

Another fact worth to be discussed is the capture of disc material by
the perturbing star. As mentioned in Section \ref{sec:results} about
0.003~\tmsun~=~3~\tmjup\ are accreted by the perturber in a typical 500~AU
encounter like model E/X002, while the amount of accreted gas
can be as large as 10~\tmjup, as in models E/X003
(see Table \ref{tab:sims}). An even larger amount can be contained in
a circumstellar disc or envelope around the perturber.
Although being smaller than
typically assumed for the Minimum-Mass Solar Nebular (MMSN, see
\citealt{Crida09}) this may still be sufficient for the formation of
Jupiter- or Saturn-type planets around the perturber. Since the orientation
of this encounter-related accretion disc is not correlated to the
stellar rotation of the perturber, this scenario may provide an explanation
for highly inclined or even retrograde planets which have recently
been detected \citep{Naritaetal09,Johnsonetal09}. If the captured gas is
accreted onto a pre-existing circumstellar disc of the perturber star
this might even lead to the formation of planetary systems with multiple
mutually inclined (or even retrograde) orbital planes.

We have to
emphasize at this point that gas capture from a massive extended circumstellar
disc is only one of several possible gas capture scenarios, and is being
observed in our work as a spin-off besides the main topic of this article.
As the more general case, capture of material from any dense gas aggregate
in the hosting star-forming region after the formation of the protostar itself
may be a possible channel to form non-aligned planets.
As the most general formulation, we note that the whole process of planet
formation from the pre-stellar cloud in the context of stellar encounters
in young star clusters to a fully established planetary system
appears to be a discontinuous one in many cases, probably limiting the probability
of regularly-shaped planetary systems with Solar system-like architecture.
This issue will be investigated in detail in future work.

One issue not treated by our calculations is the varying protostar and disc
mass during the accretion process. According to \citet{MIM10} the growing
disc may become temporarily unstable when the mass of the protostar is
still negligibly small ($\sim 10^{-3}\,\msun$).
In their nested-grid calculations, the disc fragments
in the region of $<100$~AU, and even within 10~AU. However, their
simulations do not include a realistic treatment of radiative transfer.
It is subject to future
investigations whether accreting discs under perturbation may develop
into planetary systems with Solar-type architecture.

Furthermore, the influence of magnetic fields are completely ignored
in our calculations. \citet{LMU10} show that both hydromagnetic and
thermomagnetic effects may amplify density waves in the disc into
instability, and may provide an effective viscosity via turbulence
which may assist accretion onto the star.

Observations of tidally perturbed fragmenting discs would surely be
the best confirmation of this scenario. Due to the short duration
of the flyby of only about 10,000 years the chance of such an event to
be 'caught in the act' is small. It may, however, be
possible to identify encounter-triggered fragmentation if observers succeed
in detecting the typical tidal arms around the disc-hosting star as
well as smaller amounts of non-circular filaments around a close-by
star indicating it as a candidate perturber star. Such structures may
be detectable with future high-resolution instruments like ALMA in
star-forming regions like the ONC.  Another possibility is to target
FU Orionis stars which are thought to be undergoing enhanced
accretion, such enhanced accretion may be caused by the perturbation
of discs by a close encounter.

\section{Summary and Conclusions}\label{sec:conclusions}
In a series of SPH calculations we have shown that massive ($\sim0,5\,\msun$),
extended ($\ge100\unit{AU}$) circumstellar discs can
be stable when isolated, but fragment when perturbed by a moderately close
and slightly inclined stellar encounter.
Binaries formed in two cases via
triple encounters of companions and via grazing encounters of accreting
envelopes. This agrees with binary formation in self-fragmenting discs
\citep{StaWhi09a}. We further found that the mass distribution of
companions formed in discs is in agreement with the canonical substellar
IMF as a separate population \citep{Kr01,KB03b,TK07,TK08}.
We have, however, to note that
direct tidally induced fragmentation is probably only responsible for a
fraction of all disc fragmentation events due to the relatively small
orbital parameter subset that is suitable for fragmentation of discs
of about 0.5~\tmsun. More massive discs are expected to be more prone to
fragmentation even due to weak perturbations, but are also more likely
to reach the limit for self-fragmentation. This study, however, shows that
tidal perturbations do not necessarily inhibit fragmentation but are
instead capable of inducing fragmentation in disc that otherwise would
silently disperse or be accreted by the central star without ever
experiencing fragmentation. The perturber star may accrete gas from the
target circumstellar disc and may form planets on misaligned or
even counter-rotating orbits with respect to the stellar spin.
Future work will analyse discs with different
masses and more or less massive host stars as well.

\section*{Acknowledgements}
This work was partially supported by the University of Bonn
and partially by DFG grant
KR1635/12-1.
IT wishes to thank Dr. Sambaran Banerjee and Dr. Ole Marggraf for their
kind assistance in improving this paper.\\ %clear stuck text between text body and bibliography

\end{document}